\documentclass[10pt,conference]{IEEEtran}
\usepackage[utf8]{inputenc}
\usepackage{cite}

\begin{document}

\title{What Makes Research Software Sustainable?
An Interview Study With Research Software Engineers.}

\author{
  \IEEEauthorblockN{
    M\'ario Rosado de Souza\IEEEauthorrefmark{1},
    Robert Haines\IEEEauthorrefmark{2},
    Markel Vigo\IEEEauthorrefmark{2},
    Caroline Jay\IEEEauthorrefmark{2}
  }
  \IEEEauthorblockA{\IEEEauthorrefmark{1}University of Lavras, Lavras, 37200000, Brazil}
  \IEEEauthorblockA{
    \IEEEauthorrefmark{2}University of Manchester, Manchester, M13 9PL, UK\\
    Email: caroline.jay@manchester.ac.uk
  }
}

\maketitle

\begin{abstract}
Software is now a vital scientific instrument, providing the tools for data collection and analysis across disciplines from bioinformatics and computational physics, to the humanities. The software used in research is often home-grown and bespoke: it is constructed for a particular project, and rarely maintained beyond this, leading to rapid decay, and frequent `reinvention of the wheel'. Understanding how to develop sustainable research software, such that it is suitable for future reuse, is therefore of interest to both researchers and funders, but how to achieve this remains an open question. Here we report the results of an interview study examining how research software engineers -- the people actively developing software in an academic research environment -- subjectively define software sustainability. Thematic analysis of the data reveals two interacting dimensions: \emph{intrinsic sustainability}, which relates to internal qualities of software, such as modularity, encapsulation and testability, and \emph{extrinsic sustainability}, concerning cultural and organisational factors, including how software is resourced, supported and shared. Research software engineers believe an increased focus on quality and discoverability are key factors in increasing the sustainability of academic research software.
\end{abstract}

\section{Introduction}

``Software turns a theoretical model into quantitative predictions; software controls an experiment; and software extracts from raw data evidence supporting or rejecting a theory'' (p.1)~\cite{Pradal2013}. This statement highlights the central role of software in modern scientific discovery. Significant effort goes into the development of research software, and substantial resources (either human, financial, administrative or infrastructural) are devoted to ensuring its success. Research software has some peculiarities that make it different from enterprise software: it is built in a collaborative fashion by individuals who have temporarily aligned interests (i.e.\ specialists in a field teaming up with software engineers)~\cite{Hong2008}; and it is developed using time-limited resources --- such as a research grant --- placing severe constraints on its lifespan and threatening rapid obsolescence~\cite{Seacord:2003}. Software decays relatively quickly if it is not maintained and this is especially true for software used in research~\cite{Hong2008}.

Research into what makes software sustainable is a growing field~\cite{Calero:2013, Venters:2014, Aljarallah:2018}. Software sustainability covers a broad range of concepts, related to both environmental sustainability~\cite{Calero:2013}, and the longevity of a codebase~\cite{Aldabjan2016, Alhozaimy2017}. In this paper we consider only the latter, within the context of research, where it is of particular importance given the centrality of software to the scientific process~\cite{Crouch:2013}. The issue of software sustainability is particularly relevant to ensuring a rigorous application of the scientific method in general, and to guaranteeing the fundamental principles of comparability, replicability and reproducibility in particular, which are at risk if software is not fully accessible and functional. In order to preserve these qualities, we must understand how to build software to last beyond the time-frame determined by the duration of a project, and how to increase its visibility, accessibility and findability so that it continues to be used, tested and extended by others.

Surveys of software engineers in industry have shown that software characteristics such as security, usability, reliability and maintainability~\cite{Aldabjan2016} and functional correctness, availability, and interoperability~\cite{CONDORIFERNANDEZ2018289} are considered important for sustainability. Sustainability remains a relatively nebulous concept, however, with few software engineers demonstrating a solid grasp of what it entails~\cite{Groher2017}.

Here we examine sustainability from the perspective of research software produced in an academic environment, interviewing research software engineers (RSEs) to determine whether there is a shared understanding of what sustainability means. Our results suggest that whilst there is some consensus as to the general meaning of the concept, there are a variety of views about the best way of achieving it. In particular, RSEs recommend paying attention to software quality --- an issue identified for improving the sustainability of all types of software --- as well as actions more specific to research software, including improving discoverability through building a community around a project, and raising awareness of the importance of software curation.

\section{Methods}
\label{sec:methods}

Data was collected over two phases. During phase one, interviews were conducted with nine developers at a single UK university. The preliminary results from this study were reported in a short non-archival workshop paper~\cite{Souza2014}. Here we combine this data with interviews conducted with ten further developers from four institutions: the original university, two additional UK universities and a UK Government-funded research facility.

Altogether, nineteen research software engineers (3 female, 16 male) were recruited through purposive and snowball sampling. The participants had worked on a variety of projects within different research groups and had between 18 months and 20 years of software engineering experience.

The semi-structured interviews, which were conducted either face-to-face or via video-conferencing, used the following schedule (probing questions in italics): From your point of view, what is sustainability in terms of software? \emph{(What are the attributes or features of the software that lead you to believe that it is sustainable?)}; Regarding the software you've developed: was sustainability a consideration? \emph{(If yes, at what point in time did it become a consideration? If no, why not?)}; Have you worked on any projects that were not sustainable? \emph{(Were there any consequences of it not being sustainable?)}.

Prior to each interview, participants were provided with an information sheet, and written informed consent was obtained. The mean interview time was nine minutes and thirty three seconds. The interviews were recorded, transcribed, and uploaded to the qualitative data analysis software, Dedoose 4.122. All interviews were treated as a single dataset, i.e., we did not distinguish between the two collection phases as there was no methodological reason to do so, and the numbers were sufficiently small that comparison between institutions would not have been appropriate. Transcripts were thematically analysed in an open coding fashion following established analysis methods: (1) familiarisation with the data; (2) generating the initial codes; (3) searching for themes; and (4) iteratively reviewing themes~\cite{Braun2006}. The complete dataset was coded by two researchers independently (one of whom did not participate in the study design). Inter-coder agreement was measured by computing Cohen's Kappa for whether the coders both noted that a theme was reported by a participant. A coefficient of 0.82 indicated substantial agreement. Disagreements were resolved via discussion. The full dataset is archived on Zenodo~\cite{Souza2019_data}.

\section{Results}

 Section~\ref{sustainablity_definition} summarises participants' views on what sustainability encompasses with illustrative quotations, where the code in brackets after each quotation indicates the participant identifier and study phase in which the data was collected (for example, P3-S1 refers to participant 3, study phase 1). The results of a thematic analysis, which classifies sustainability concerns as \emph{intrinsic} or \emph{extrinsic} are reported in Section~\ref{sustainability_features}.

\subsection{Conceptual Understanding of Sustainability}
\label{sustainablity_definition}
Eighteen participants recognised sustainable software as that that was reusable, either in its original project, or another project in the future: `it means someone should be able to build, run, and understand your software, say one year, five years, or ten years after you're finished with it, and without having to come and ask you how to build it, or why you did things in a certain way' (P9-S2); `sustainability is basically making sure that software works overtime. So even if development stops on a particular software product, a couple of years down the line you can still download it and it would work' (P7-S1). One participant had not heard the term used in relation to software before, but was familiar with the concept when it had been explained, and recognised it as desirable.

Nine participants said unequivocally that sustainability was a consideration: `Yeah always has been. We try and keep the software sustainable in the hope that either it will get more funding to continue it, or if we don't get funding it would be a shame to see it all die and disappear so it would be nice if it was in a state that someone else could pick it up and use it' (P9-S1). Two participants said sustainability was sometimes a consideration, and one was not sure. Six participants said it was becoming more important, and three of these identified their current projects coming to an end as a reason for this: `I don't think it was a primary consideration, but it is becoming more important now that a lot of our projects are coming to an end, and we need to make a plan for them to be maintained in the future' (P5-S1). One said he had not thought about it consciously, but that his manager may have done.

Ten participants said they considered sustainability at the beginning of a project, and seven said it was considered after some time: `I don't think it was a consideration at the start, I think at the start it was more about getting things done, getting things ready, so yeah it's more of a thing that's come about as the project has come along' (P3-S1). `I've worked here for quite a few years so I'm used to this whole funding cycle and, you know you're paid for three years and then after two and a half years things start getting a bit hairy and you're hoping for more funding, so you know, with that in mind, we try and keep the software sustainable' (P2-S1).

Fourteen participants reported that they had worked on software that was not sustainable. The remaining six said they did not think so, or it was hard to say, although it should be noted that they were not claiming that all software they had produced still worked, but rather that they considered the software to be sustainable when they stopped working on it.

\subsection{Features of sustainability}
\label{sustainability_features}

A thematic analysis of the data indicates that sustainability must be considered from two perspectives: \emph{intrinsic sustainability} and \emph{extrinsic sustainability}.

The \textbf{\emph{intrinsic sustainability}} of software concerns characteristics of the software artefact itself, and includes factors relating to how the code is written and documented. The themes that emerged for intrinsic sustainability are described below. Table~\ref{tab:frequencies} lists each theme and the frequency with which it was mentioned. It should be noted that whilst the themes have been reported as distinct, they are often interrelated.

\subsubsection*{Documented}
Participants largely agreed that code must be well documented for sustainability to be possible: `\ldots there are some additional steps that you have to do, like you make sure you have documentation, you make sure that the source code is in one place' (P8-S1); `It needs to be well-documented' (P5-S1). `I remember spending a couple of months writing documents, analysing every single module and itemising [\ldots], doing reverse documentation [\ldots] just to demonstrate to the management whether this software was usable or not' (P4-S2).

\subsubsection*{Testable}
Several developers stated that testing is important: `It's a lot of test automation and continuous integration testing, and I think that helps a lot with keeping it sustainable' (P2-S1); `Software tests as well. Yes, absolutely' (P7-S1).

\subsubsection*{Readable}
There was a general belief that if code is easy to read it will be more sustainable, because it will be more straightforward for someone else to pick up: `\ldots if he (someone other than the original developer) finds my code, and found that the effort of learning to use my code is going to be more difficult than the actual benefit it gave him, he'd probably throw it away and write his own stuff' (P1-S1).

\subsubsection*{Modular}
Breaking software up into component parts with well defined interfaces was viewed as making it easier for others to reuse the software as a whole, or a subset of its constituent parts: `It turned out that the software was impossible for anyone to actually deploy in full, and it would only work if all the pieces were deployed. Funny, that didn't work' (P4-S1). `People don't appreciate how encapsulation is really a good sustainable practice because it means things are more understandable, you know for somebody who's new to the software' (P8-S2).

\subsubsection*{Standardized}
Two developers made it clear that ``reinventing the wheel'' should be avoided, particularly when support is often good for libraries that have a large user base. `It's important to [use] technologies that people generally understand, reusing as much as you can, so don't write your own things, [when] there's good solutions already' (P6-S1).

\subsubsection*{Useful}
If the software is fulfilling its purpose in an effective way, people will be motivated to sustain it. `\ldots it's coupled to the software doing something useful, which either there isn't an alternative for, or that it is much better in its niche than the alternatives' (P4-S1). Alternatively, if the code does not fulfill the purpose precisely, `they think, ``OK, I will take the idea, but I will write my own stuff'''(P3-S1).

\subsubsection*{Scalable}
Making code scalable was thought to help future-proof it. This ensures `\ldots it's also going to be usable long term, because if it's just the simple cases, people go, ``yeah that's a really nice idea'', and then as soon as they start using it in anger, a lot blows up because it doesn't scale' (P1-S1). `the cost of adding new features should not increase exponentially, as it does in some codes' (P1-S2).

Whilst intrinsic sustainability concerns the application code, \textbf{\emph{extrinsic sustainability}} concerns the environment in which the software is developed and/or used. Extrinsic factors can be separated into the following broad, interrelated themes.

\subsubsection*{Openly available}
Sharing research software in an open repository after the project ends increases the chance it will be found and reused. `Usually I would look online in a repository for libraries and I would see when it was last updated \ldots if it's in version control then it's a good start' (P9-S1).

\subsubsection*{Shared/co-owned}
If the software is developed by a team, this increases the chances of it remaining active. `[It's important] that there is some community around it. You need to have more than one person involved, right? If it's a one man project and that guy is hit by a bus or just decided to do something else, work at Google or something, then it just dies' (P8-S1). `Whether that community is composed of volunteers or people that are actually paying for your product it doesn't matter. But basically you do need to have a community, or at least you have to have a very dedicated individual' (P3-S2).

\subsubsection*{Resourced}
This is one of the aspects that developers were most concerned about. `\ldots a lot of our projects are coming to an end and we need to make a plan for them to be maintained in the future' (P3-S1). `You see it in a lot of research, I mean the [removed for anonymity] stuff I did -- [it's] completely just gone. The minute I left it, still sitting on GitHub but no one even looked at it' (P1-S1).

\subsubsection*{Actively maintained}
Developers are wary of software not in current usage, due to the potential for out-of-date dependencies and modules that no longer work because the platform has evolved. `Physically the software lies there \ldots you find software to do something, [you think] OK that looks good, and then you look -- last updated three years ago. Most people won't touch it' (P1-S1). `So it's about having this kind of momentum to the project, so that it keeps moving. That you have further development, even if you have maintenance mode---that is, not many new buttons being added---but at least there is someone [keeping] it alive (P2-S1). `I guess it is around sort of maintainability, the fact that codebases, if you don't sort of keep them up-to-date and keep developing them, they tend to go stale' (P6-S2).

\subsubsection*{Independence from infrastructure}
Sustainability can be related to where the software runs; if the infrastructure is not maintained, is the software capable of running outside that environment? `[Removed for anonymity] did most of that, and it's one of these things that will probably stay alive for as long as the server that it's on stays alive, and if that server crashes they will probably not bother rebuilding it into another machine' (P1-S1). `One would be that the code should run on hardware for the next foreseeable future. So that means, that it's sufficiently portable' (P1-S2).

\subsubsection*{Supported}
Sustainable software usually has some sort of user facing support from the team who is developing or maintaining it, which is helpful to both external developers and end-users. This is directly related to the project being active: `So this tends to mean things like e-support, or automated tools of various kinds' (P4-S1).

\subsubsection*{Version-Control}
Managing changes to source code in an organized fashion is a desirable feature of software that is sustainable. This enables developers to restore previous versions of the software as well as understanding what happened to the code over time and tracking bugs. `I would say sustainability is partly about you know this use of re-usability of research \ldots to keep track of, your developments because you want to see exactly which version of the code produced what results that went into some application' (P8-S2).

\begin{table}
  \centering
  \caption{Frequencies of Themes Per Study Phase}
  \label{tab:frequencies}
  \begin{tabular}{rccc}
    \hline\hline

    & Phase 1 & Phase 2 & Total \\

    \hline
    \textbf{Intrinsic sustainability} & & & \\
    \hline

    Documented   & 8 & 6 & 14 \\
    Testable     & 5 & 2 & 7  \\
    Readable     & 5 & 7 & 12 \\
    Modular      & 2 & 3 & 5  \\
    Standardized & 2 & - & 2  \\
    Useful       & 2 & 2 & 4  \\
    Scalable     & 1 & 1 & 2  \\

    \hline
    \textbf{Extrinsic sustainability} & & & \\
    \hline

    Openly available              & 6 & 6 & 12 \\
    Shared/co-owned               & 3 & 4 & 7  \\
    Resourced                     & 6 & 4 & 10 \\
    Actively maintained           & 6 & 7 & 13 \\
    Independent of infrastructure & 1 & 6 & 7  \\
    Supported                     & 4 & - & 4  \\
    Version-controlled            & - & 4 & 4  \\

    \hline\hline

  \end{tabular}
\end{table}

\section{Discussion}

The results of the study indicate that the Software Sustainability Institute's view that sustainable software `will be available---and continue to be improved and supported---in the future' is, understood by developers, and considered to be meaningful. It is not clear that the working definition used by most developers is an exact match, however. To end users, the Software Sustainability Institute's definition essentially means a software application that they can continue to use, but the research software engineers in our study considered sustainability to be a much broader concept. In particular, there was a significant focus for many developers on trying to ensure some aspect of the code itself is usable in the future, regardless of whether that use occurs in the same application, or contributes to a different one.

Although there is no concrete guidance on how to achieve sustainability in research software engineering~\cite{Raturi:2014}, many of the factors that developers consider to be important for sustaining research software have also been noted as important for sustaining enterprise software products~\cite{Seacord:2003} and for successful Free/Libre Open Source Software (FLOSS) development~\cite{Crowston:2008}. We therefore suggest that from the perspective of research software, a broad view is helpful: in simple terms sustainability can be considered at the level of a software product delivered by a particular project. For example, a project could provide software with a certain amount of functionality and a sustainability specification along with it.

Other works converge on the themes we have identified as contributing to more sustainable software. Chue Hong \& Voss (2008) and Seacord et al. (2003) suggest that building an ``ecosystem'' around the software can generate an environment where it can thrive~\cite{Hong2008,Seacord:2003}. This points to the necessity of paying attention to extrinsic factors affecting the software, as well as the social context in which it is developed.

 Seacord et al. (2003) highlights the importance of the modernization and change potential of a software application. This potential is the differential between the actual properties of the system and the desired properties~\cite{Seacord:2003}. Even if attention is paid to intrinsic aspects of the artefact, these cannot provide a measure of sustainability alone. The amount of time a team (or individual) would need to spend to bridge that gap must also be taken into account.

Calero et al. (2013) proposes four characteristics of software and digital objects within the e-Research environment which help them to ultimately be sustainable, and can be mapped to our dimensions of sustainability as detailed below \cite{Calero:2013}:

\begin{enumerate}
\item Understanding of the requirements of the current and potential users of the object \emph{(useful)}
\item Improvement of the object to increase the potential number of users \emph{(supported, actively maintained)}
\item Identification and dissemination of research outputs which have resulted from the use of the object \emph{(openly available)}
\item Increase in the community involvement around the object \emph{(shared/co-owned)}
\end{enumerate}

Following good development practices is a concept widely identified in the literature in the form of system architecture, design documentation and test scripts~\cite{Seacord:2003}. In the absence of infinite resources, however, projects --- and the software they produce --- are going to remain of a fixed term nature. In this case, the route to sustainability is likely to be via software reuse in future projects.

\subsection*{Acknowledgements}
This work was funded by Brazil's Science without Borders programme, and EP/N006410/1, EP/S021779/1.

\bibliographystyle{ieeetr}
\bibliography{sustainability}

\end{document}